\documentclass{appolb}
\usepackage{graphicx}
\usepackage[OT4]{fontenc}
\usepackage{cite}
\pdfoutput=1
\newcommand{\liliput}{LILLYPUT~3}
\begin{document}
\title{Development of a dedicated beam forming system for material and bioscience research with high intensity, small field electron beam of {\liliput} accelerator at Wroclaw Technology Park
\thanks{Presented at Jagiellonian Symposium of Fundamental and Applied Subatomic Physics, Krak{\'o}w, 2015}
}
%
\author{Przemys{\l}aw Adrich\thanks{przemyslaw.adrich@ncbj.gov.pl}, Arkadiusz Zaj\k{a}c
\address{National Centre for Nuclear Research, {\'S}wierk, Poland}
\\
Piotr Wilk
\address{Wroclaw Technology Park, Wroc{\l}aw, Poland}
\\
Maciej Chorowski, Jaros{\l}aw Poli{\'n}ski, Piotr Bogdan
\address{Wroc{\l}aw University of Technology, Wroc{\l}aw, Poland}
}
\maketitle
\begin{abstract}
The primary use of the {\liliput} accelerator at the Nondestructive Testing Laboratory at Wroclaw Technology Park is \mbox{X-ray} radiography for nondestructive testing, including R\&D of novel techniques for industrial and medical imaging.
The scope of possible applications could be greatly extended by providing a system for irradiation with electron beam. The purpose of this work was to design such a system, especially for high dose rate, small field irradiations under cryogenic conditions for material and bioscience research.
In this work, two possible solutions, based either on beam scanning or scattering and collimation, were studied and compared. It was found that under existing conditions efficiency of both systems would be comparable. The latter one was adopted due to its simplicity and much lower cost. The system design was optimized by means of detailed Monte Carlo modeling. The system is being currently fabricated at National Centre for Nuclear Research in {\'S}wierk.
\end{abstract}
\section{Introduction and motivation}
\begin{figure}[!hbt]
  \centering
	\includegraphics[width=0.8\columnwidth]{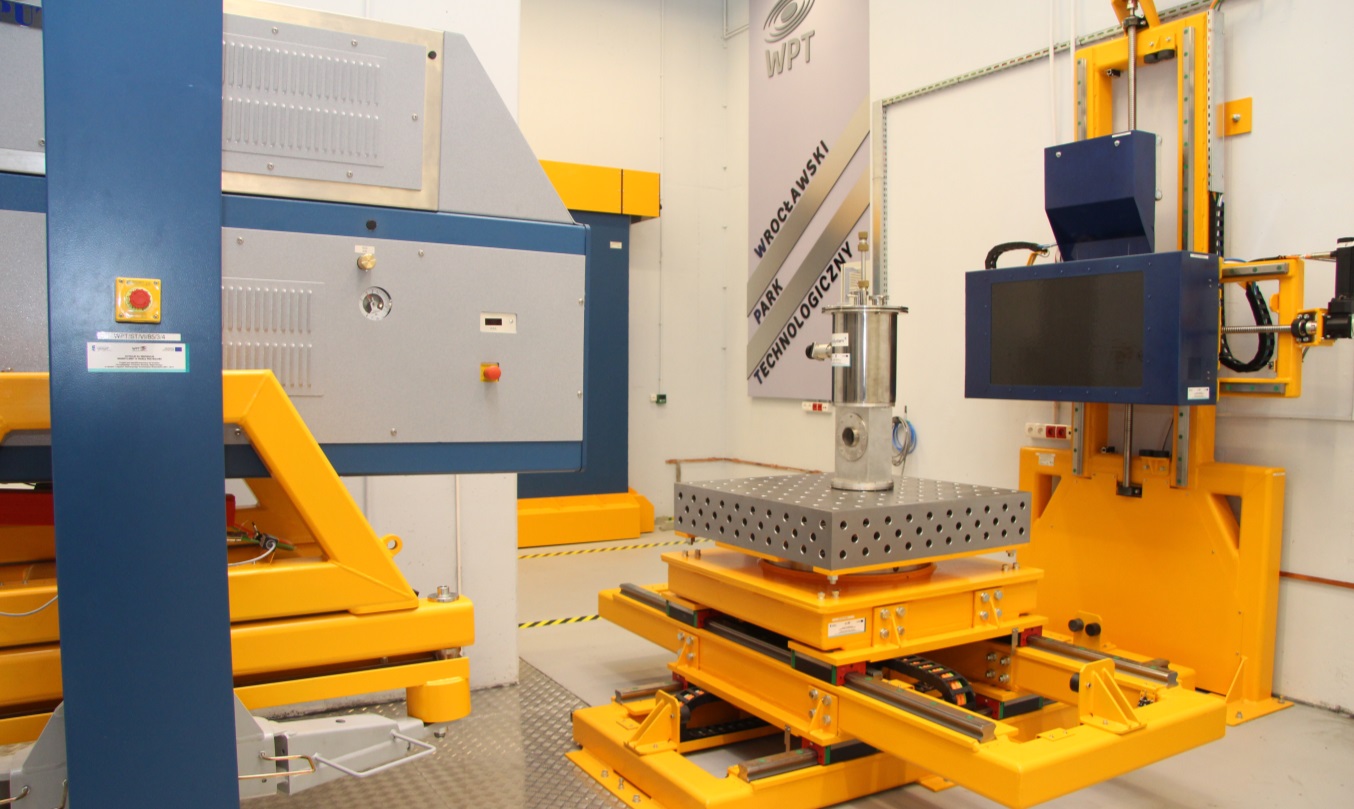}
  \caption{\small The {\liliput} system in a standard radiographic configuration installed at Nondestructive Testing Laboratory at Wroclaw Technology Park. The accelerator is visible on the left, the object table in the center and digital imaging system to the right.}
  \label{fig:Lillyput3}
\end{figure}

The {\liliput} accelerator belongs to the line of radiographic electron linear accelerators developed and produced commercially by the National Centre for Nuclear Research (NCBJ). Recently an accelerator of this kind was delivered, installed and commissioned at the Nondestructive Testing Laboratory (NDT) at Wroclaw Technology Park (WPT). In a standard, radiographic configuration {\liliput} delivers a selectable 6 or 9~MV \mbox{X-ray} beams. The system delivered to Wroclaw is complete with a digital imaging system and an automatized object table, both designed and produced at NCBJ. The complete system is presented in Fig.~\ref{fig:Lillyput3}.

The functionality of the accelerator, and thus the scope of its application and overall utilization of the NDT laboratory, could be enhanced by providing a system for irradiation with electron beam. Such a system is not part of the standard configuration of the {\liliput}, although the construction of the accelerator can, with minor modifications, accommodate it.
It is the purpose of this work to design a system for electron beam irradiation for the {\liliput} accelerator at NDT.
This development is specifically motivated by research planned at NDT in the area of novel insulator materials capable of withstanding absorbed doses of radiation in the range of tens of MGy. 
The secondary motivation for this work includes R\&D into sterilization of novel, polymer based medical products and studies of structure modifications of irradiated polymers. For polymer irradiations, absorbed doses in the range of tens of kGy are sufficient.
\section{Purpose of the beam forming system}
\begin{figure}[!hbt]
  \centering
	\includegraphics[width=0.9\columnwidth]{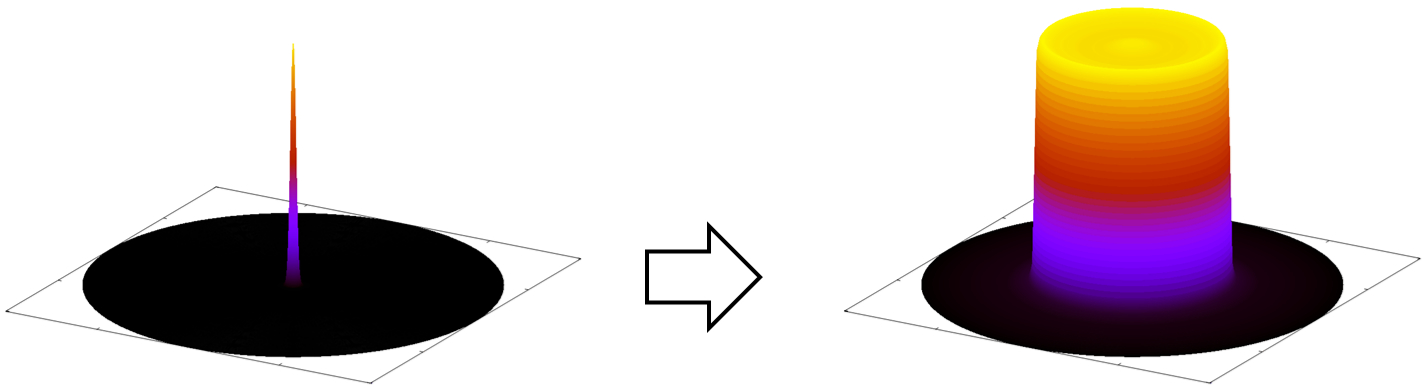}
  \caption{\small Schematic comparison of electron beam fluence distribution of the primary ``pencil'' beam with FWHM=3~mm, accelerated in an electron accelerator (to the left) and a desired beam at an irradiation plane (to the right) with FWHM=80~mm. The purpose of the beam forming system is to transform the former into the latter.}
  \label{fig:beam_forming_purpose}
\end{figure}

Apart from the source of energetic electron beam, i.e. the accelerator, a key component of an electron irradiation system is the beam forming system. As schematically illustrated in Fig.~\ref{fig:beam_forming_purpose}, the purpose of the electron beam forming system is to transform the primary ``pencil'' beam delivered by a linear electron accelerator with typically a Gaussian profile of FWHM$\approx2-3$~mm, into a wide (FWHM on the order of centimeters to tens of centimeters) and uniformly distributed beam as required in any practical application.
\section{Specific requirements and constraints on the electron beam forming system at NDT}
\label{sec:requirements}
\begin{figure}[!hbt]
  \centering
	\includegraphics[width=0.5\columnwidth]{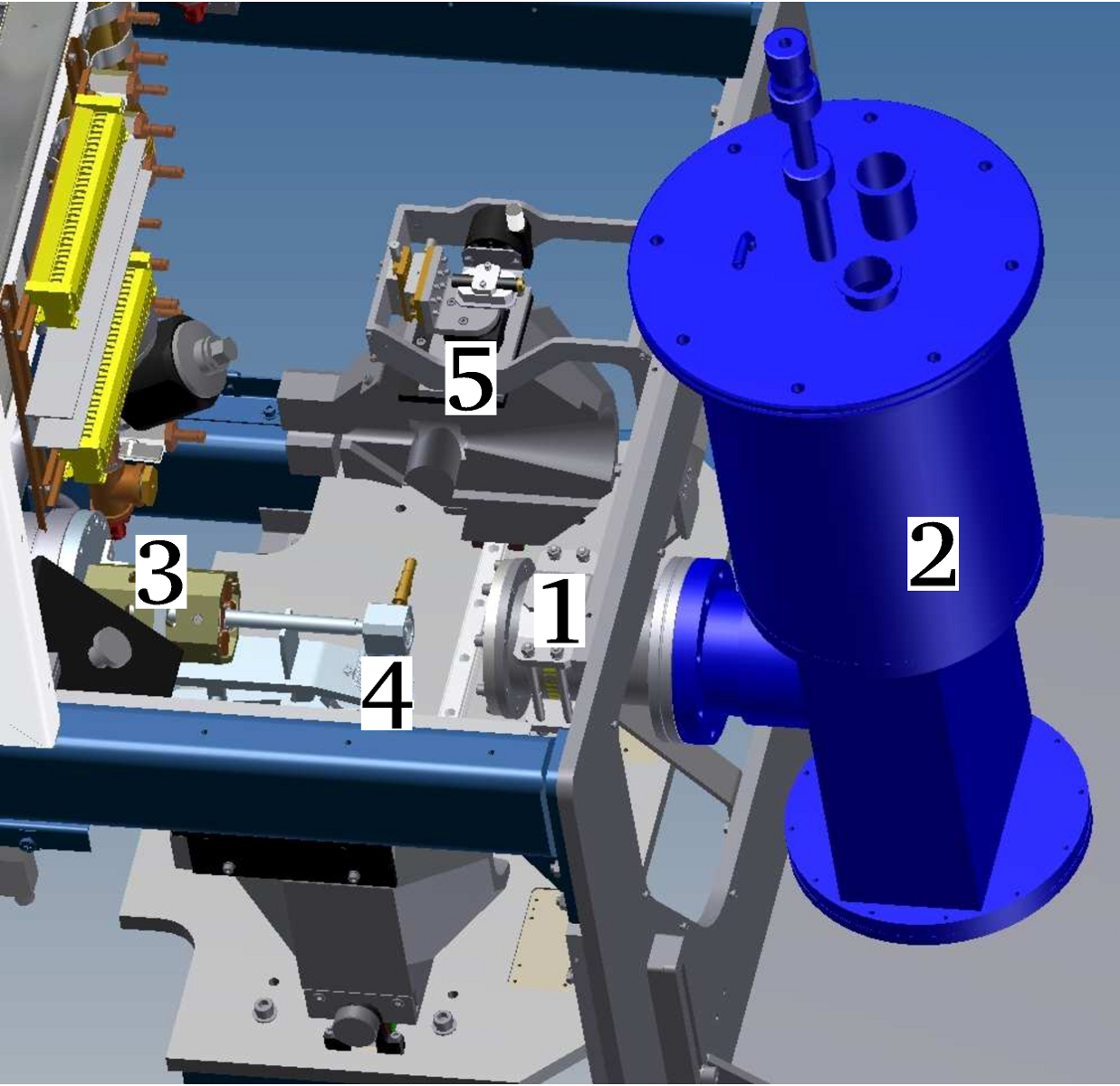}
  \caption{\small 3D visualization of an applicator tube (labeled ``1'') attached to the LN$_2$ cryostat (``2'') and part of the {\liliput} accelerator: beam focusing coils (``3'') and target chamber with titanium beam exit window (``4''). Also visible is an X-ray beam collimator (``5''), moved to the side of the beam axis to make space for the electron beam forming system.}
  \label{fig:cryostat_applicator_3d}
\end{figure}

The research planned at NDT requires delivery of 9~MeV electron beam with as high as possible dose rate at the irradiation plane and at the same time flattened to within few percent over a field of $\phi=82$~mm diameter. The samples for the irradiation are going to be kept under cryogenic conditions, in a LN$_2$ cryostat. The electron beam has to be delivered directly to the surface of a sample inside of the cryostat. This is going to be achieved by means of an applicator tube that tightly fits in an opening of the cryostat. The applicator tube has to be filled either with vacuum or with helium gas in order to avoid vapor condensation inside the tube and on a thin window separating the inside of the applicator from the inside of the cryostat.
Due to mechanical constraints, the distance between the accelerator exit window and the irradiation plane inside the cryostat cannot be shorter than 400~mm. Fig.~\ref{fig:cryostat_applicator_3d} shows a 3D visualization of the cryostat (dark blue object to the right), the electron beam applicator (gray tube connected to the cryostat from the left) and parts of the accelerator assembly, namely beam focusing coils and target chamber equipped with 50~$\mu$m thick beam exit window made of titanium foil. Dismounting of the titanium exit window, although beneficial in one of the contemplated solutions of the beam forming system, was excluded from consideration as it is a complex operation that brings risk of major disruption of the NDT operations.
%
\section{Possible solutions of the electron beam forming system}
\begin{figure}[!bth]
  \centering
	\includegraphics[width=1.0\columnwidth]{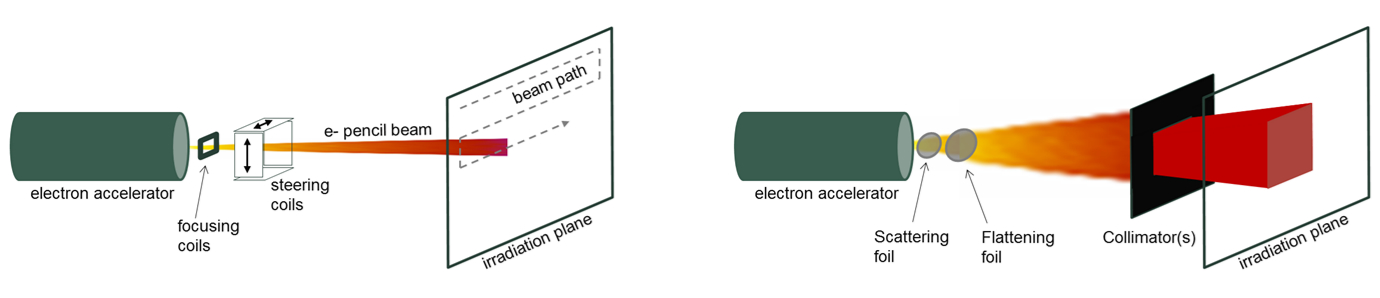}
  \caption{\small Two possible solutions of electron beam forming - a beam scanning system (left) and a scattering and collimation system (right).}
  \label{fig:activeVSpassive}
\end{figure}

The problem of forming a usable electron beam can be addressed in two different ways. The first possible solution is beam scanning, schematically depicted in the left side of Fig.~\ref{fig:activeVSpassive}. 
In this solution, the beam exiting the linac is deflected by magnetic field. The field oscillates in time in such a way that the beam spot on the irradiation plane cyclically moves over entire irradiation field, resulting in a uniform beam fluence.	 
This kind of beam forming is common in industrial applications of electron beams \cite{Scharf1989,Hamm2012}, although in these applications the beam is scanned along one axis only.
Beam scanning is known to be very efficient as the losses of the beam on the way from the accelerator to the irradiation plane could be kept minimal,  provided that the beam is transported in vacuum. On the other hand the scanning systems are relatively complex both in construction and in control of the operation and it is not trivial to achieve uniform dose distribution, especially if the beam is to be scanned in two dimensions.

The second possible solution to the problem of electron beam forming, known as a passive one, is schematically depicted in the right side of Fig.~\ref{fig:activeVSpassive}. In a passive system, the beam fluence at the irradiation plane is flattened by means of spreading the primary electron beam in a set of foils and subsequent collimation of the scattered beam to the area of the irradiation field. Due to the nature of this solution, the passive systems are generally less efficient than scanning systems as far as the beam transmission is considered. The main advantages of the passive systems are simplicity of construction and reliability of operation. In addition, in these systems it is relatively easy to achieve exceptionally well flattened beam. The passive systems are found in all modern medical electron accelerators for electron beam therapy, in many proton therapy facilities, as well as, in some research accelerators.
\section{Estimated efficiency of a scanning system under conditions at NDT Laboratory}
\label{sec:analysis_scanning}
The scanning beam system would be the most appropriate for the facility aiming at performing high dose rate irradiations. However, in case of the {\liliput}  accelerator at NDT, to fully exploit the potential of a scanning system, the accelerator exit window would have to be removed prior to the installation of the system. As discussed in Sec.~\ref{sec:requirements}, this is not possible. 

\begin{figure}[!hbt]
  \centering
	\includegraphics[width=1.0\columnwidth]{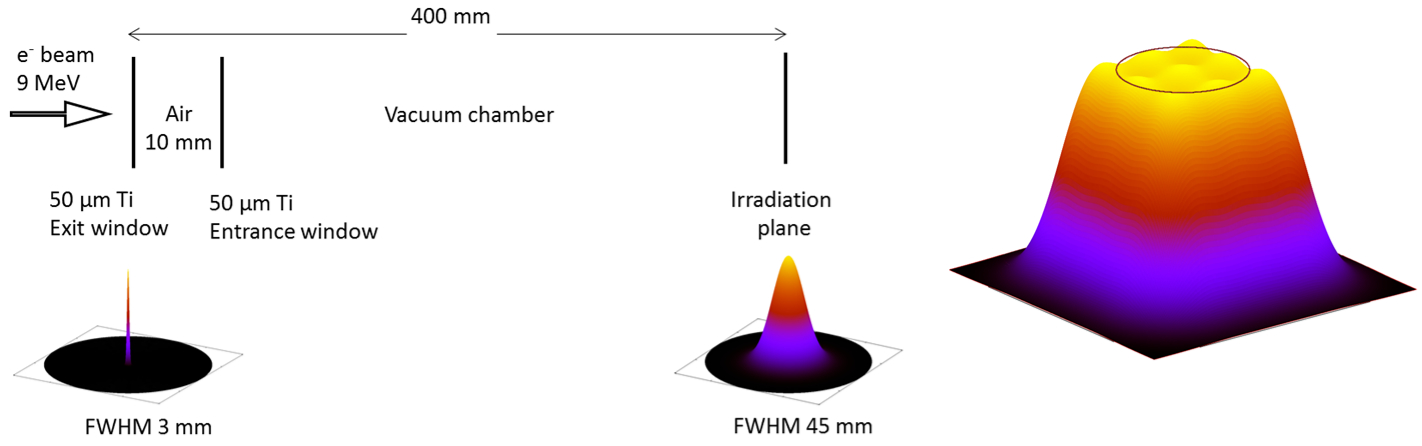}
  \caption{\small A simplified model to analyse realistic efficiency of the scanning beam solution in which the primary ``pencil'' beam has to be transported through two 50~$\mu$m thick titanium vacuum windows. Left side: a model of the system. Below the model, beam fluence distribution on exit from the accelerator (FWHM$\approx$3~mm) and on the irradiation plane (FWHM$\approx$45~mm) as calculated with a Monte Carlo method. Right side: beam fluence distribution calculated as a sum of nine Gaussian distributions each with FWHM=45~mm positioned in a 3x3 matrix, adopted as a simple approximation of beam scanning. On top of the distribution a $\phi$=82 mm field is indicated.} 
  \label{fig:scanning_reality_check}
\end{figure}

To assess a realistically achievable efficiency of a potential scanning system, we first performed a Monte Carlo calculation of beam spot broadening due to interactions with window material. A simulated system is schematically depicted in Fig.~\ref{fig:scanning_reality_check}. To minimize electron scattering in the air, an additional vacuum chamber was included. This vacuum chamber was assumed to have an entrance window identical to the exit window of the accelerator, i.e. made of 50~$\mu$m thick titanium foil. There is 1~cm space between both windows in order to allow for air cooling, what would be necessary to avoid thermal breaking of these thin windows.  As illustrated in the figure, calculations revealed, that the FWHM of the beam spot at the irradiation plane would be 45~mm. 
Reasonably flat fluence distribution can still be achieved by scanning such a broad beam. However, the efficiency of the system would be about 30\% only, as nearly 70\% of beam electrons would miss the irradiation field (as illustrated in the right side of Fig.~\ref{fig:scanning_reality_check}). A similar efficiency can be achieved in a much simpler, passive beam forming system, discussed below. 
\section{Dual-foil beam forming system}
In a passive beam forming system, schematically depicted in the right side of Fig.~\ref{fig:activeVSpassive}, the narrow ``pencil'' electron beam extracted from a linear accelerator, is first scattered in a thin, flat foil made of a high-Z material. This foil is commonly referred to as the scattering foil. A second foil, known as flattening foil, which optimally has a Gaussian radial thickness profile, $h(r) = H\exp(-r^2/R^2)$, is located at some distance, usually on the order of few centimeters, downstream from the primary foil. The parameters $H$ and $R$ depend on beam energy, field size, geometry and materials of the system. In the flattening foil, due to its variable thickness, the electrons near the beam axis (small $r$), where the foil is thickest, are most scattered, while the electrons at larger radial distance, $r$, from the beam axis are less scattered. As shown in e.g. \cite{AbouMandour1978, Grusell1994, Kainz2005, Adrich2014}, such an arrangement results in a flat fluence profile on the irradiation plane within the area of the irradiation field. Without any collimation device, electron fluence outside the field is substantial and decreases slowly with increasing distance from the beam axis. This is not desired in practical applications, therefore the beam formed in the foils is usually collimated to the area of the irradiation field, by means of an appropriate collimator and/or applicator.
%
\section{Design of the dual-foil beam forming system for the {\liliput}  accelerator at NDT}
\begin{figure}[!hbt]
  \centering
	\includegraphics[width=0.95\columnwidth]{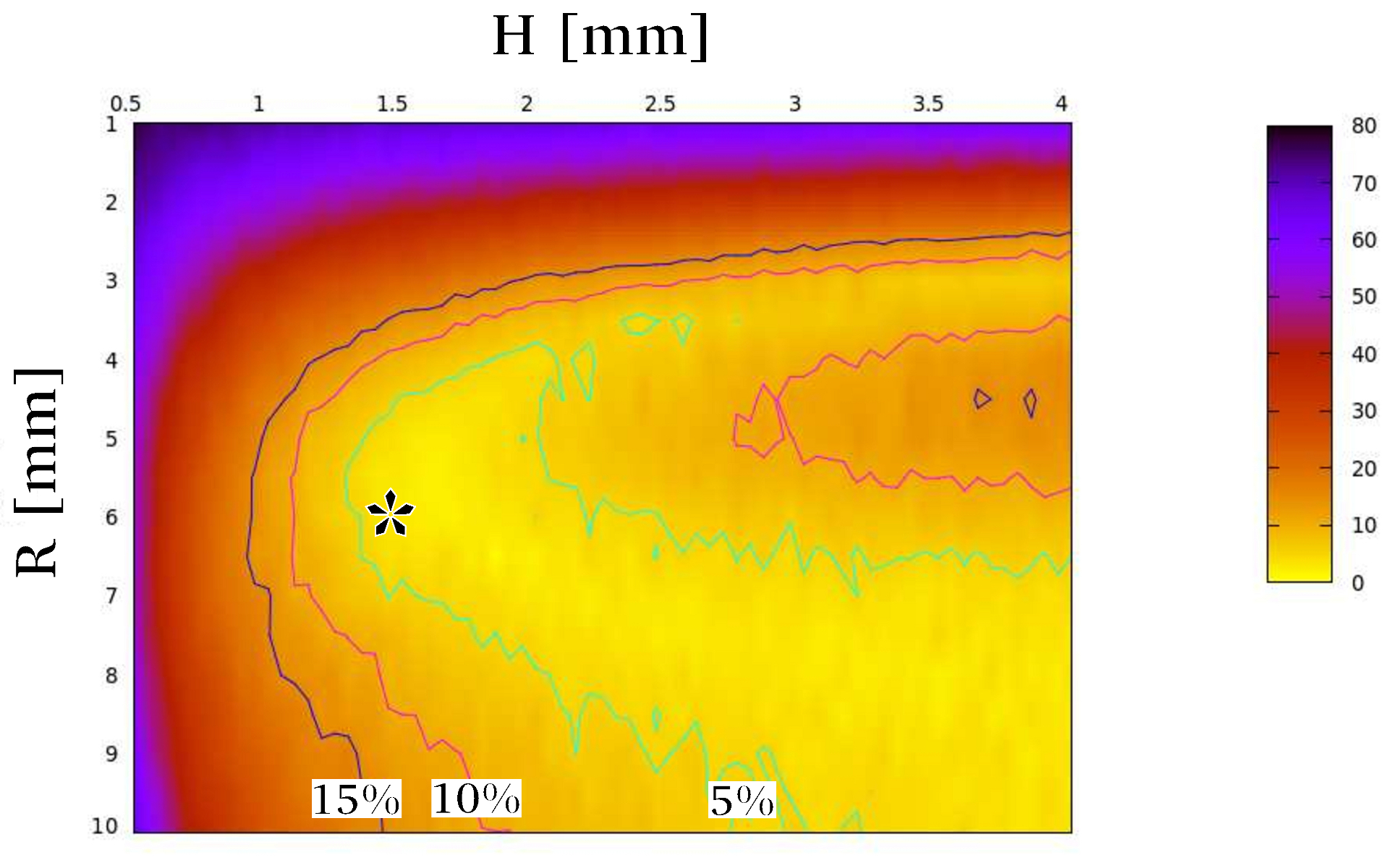}
  \caption{\small Flatness, $f(H,R)$, of the Monte Carlo calculated electron fluence distribution at the irradiation field plotted as a function of the flattening foil thickness profile parameters $H$ and $R$. Contours indicate, as labeled, areas of flatness below 5\%, 10\% and 15\%. A point of practically optimal values of $H$ (1.5~mm) and $R$ (6~mm) is indicated with a star.}
  \label{fig:HR_scan}
\end{figure}

\begin{figure}[!hbt]
  \centering
	\includegraphics[width=0.47\columnwidth]{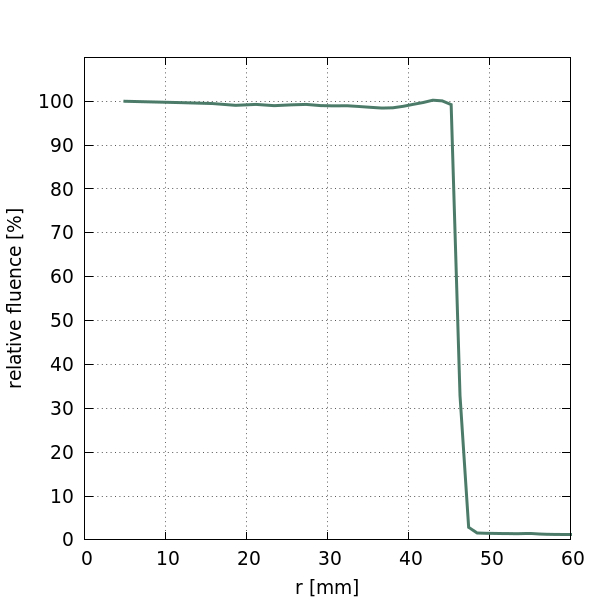}
	\includegraphics[width=0.47\columnwidth]{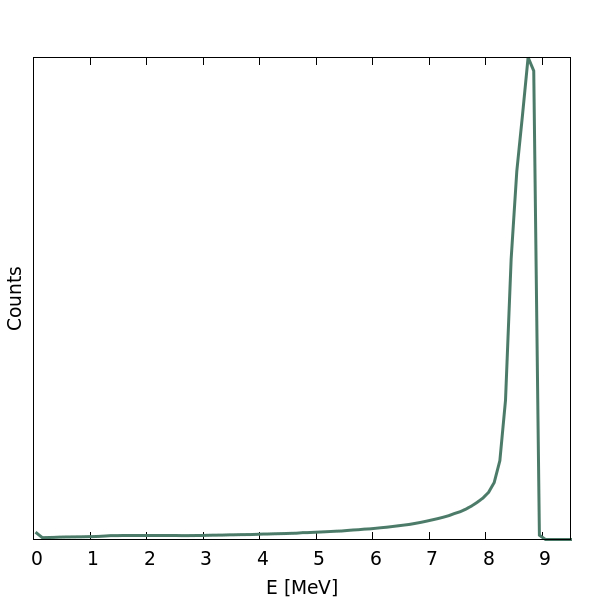}
  \caption{\small Fluence distribution and energy spectrum of electrons at the irradiation plane resulting in an optimized dual-foil beam forming system.} 
  \label{fig:Fluence_Espectrum}
\end{figure}

To design a system that would maximize the beam transport efficiency, while conforming to all requirements and constraints discussed in Sec.~\ref{sec:requirements}, a model of system geometry was constructed and its performance was assessed by means of Monte Carlo calculations of beam transport through the model. 
\begin{figure}[!bth]
  \centering
	\includegraphics[width=0.8\columnwidth]{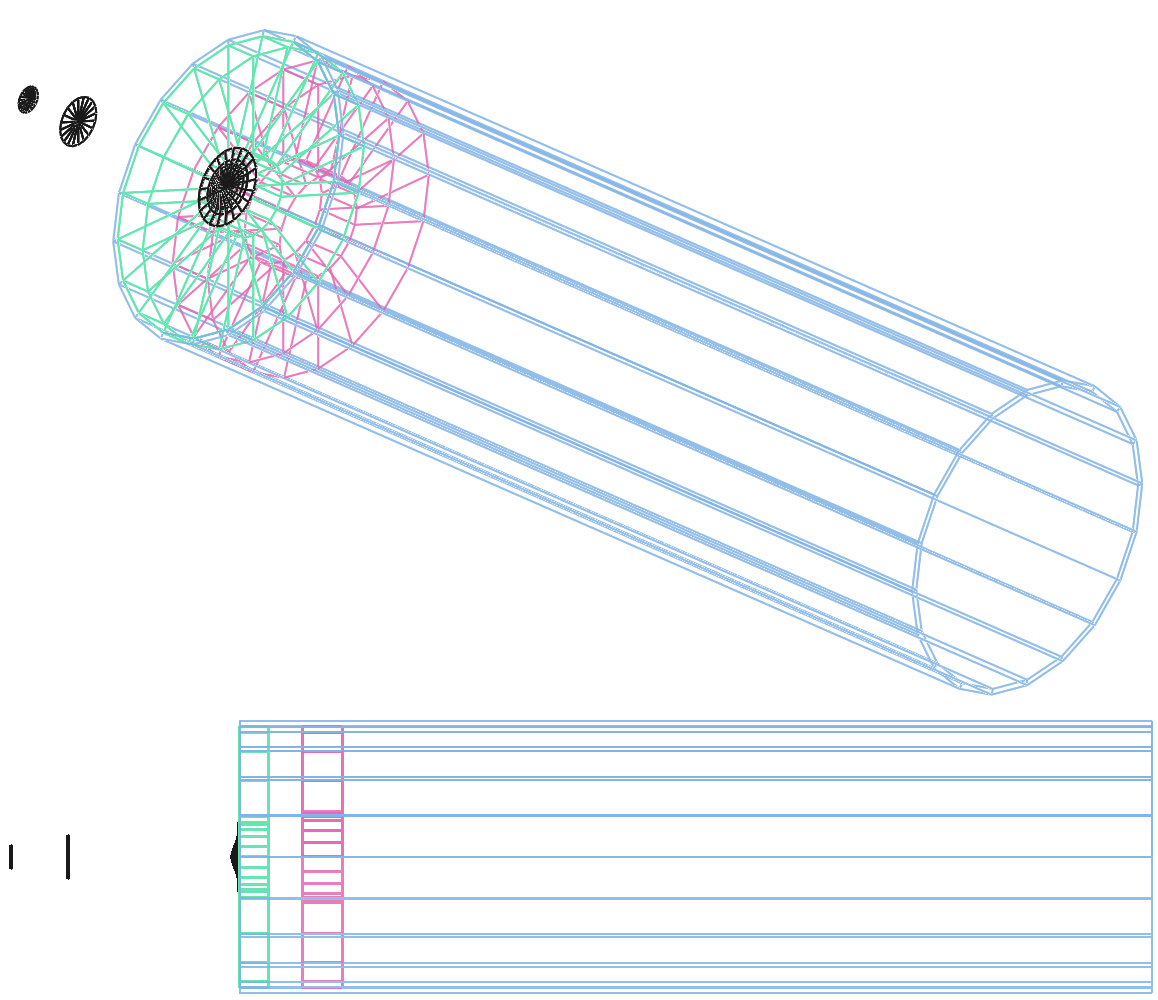}
  \caption{\small 3D visualization (top) and side view (bottom) of the Geant4 model of the system.} 
  \label{fig:Geant4_model}
\end{figure}
Fig.~\ref{fig:Geant4_model} shows a visualization of Geant4 \cite{Agostinelli2003} model of the beam forming system constructed in this work. The model includes all system components that are relevant to the forming of the beam, i.e. the components with which the beam electrons can interact on the way to the irradiation plane. Those are, in order from the left to the right in Fig.~\ref{fig:Geant4_model}, 50~$\mu$m titanium exit window, 0.01~mm tantalum scattering foil, Gaussian profiled aluminum flattening foil mounted on top of a steel flange on entrance side of the applicator tube, steel applicator tube, aluminum holder of a beam monitoring device located inside of the applicator tube. The medium between foils is air. The applicator is filled with helium gas under atmospheric pressure. As discussed in Sec.~\ref{sec:requirements}, helium atmosphere is to avoid vapor condensation inside of the applicator. This has to be accounted for in the simulation due to much lower scattering power of helium as compared to air \cite{ICRU1984}. The steel flange and an aluminum holder of the beam monitor are included because they effectively act as a beam collimator. A thin titanium exit window at the end of the applicator is not included in the simulation as it has negligible influence on the beam fluence at the irradiation plane that is right next to it.\\
Based on the results of Monte Carlo simulations all important parameters of the system are optimized. The method used for the design optimization is discussed in detail elsewhere \cite{Adrich2015}. In short, in this method, optimal values of all system parameters, besides the parameters $H$ and $R$ describing the shape of the flattening foil, are first settled based on a set of simple rules and recipes. Then, in order to establish optimal values of $H$ and $R$, i.e. values that minimize the flatness of the beam fluence distribution over the irradiation field, the $H$ and $R$ are varied in small steps over reasonable ranges. At each step, the beam fluence distribution is calculated using the Monte Carlo model mentioned above. Flatness of the calculated fluence distribution is quantified as $f = (\phi_{max}/\phi_{min} - 1)\cdot100\%$, where $\phi_{max}$ and $\phi_{min}$ are, respectively, the maximum and miniumum of the fluence distribution within the irradiation field. At the end, the flatness, $f$, calculated separately at each step, is plotted as a function, $f(H,R)$, of the parameters $H$ and $R$, as shown in Fig.~\ref{fig:HR_scan}.
Optimal flattening foil should minimize $f(H,R)$ while keeping $H$, the parameter describing thickness of the flattening foil, as small as possible in order to simultaneously minimize beam energy loss in the foil. It is evident from the plot in Fig.~\ref{fig:HR_scan}, that optimal flattening foil has $H\approx{}1.5$~mm and $R\approx{}6$~mm.\\
The electron fluence distribution, calculated at the irradiation plane for the optimized beam forming system, is shown in left panel of Fig.~\ref{fig:Fluence_Espectrum}. The beam is nearly perfectly flattened within the irradiation field. The flatness of this distribution for $r<41$~mm (i.e. over the $\phi{}=82$~mm field) is only 1.6\%. In the right panel of the same figure, the energy spectrum of electrons impinging on the irradiation plane within the irradiation field is shown. The mean of this energy spectrum, $<E>=7.9$~MeV, is one of the inputs for estimation of the dose rate (see Sec.~\ref{sec:dose_rate_estimation}). The transport efficiency, calculated as a ratio of the number of electrons registered, in the Monte Carlo calculation, within the irradiation field to the number of initial source electrons, amounts to about 39\%, thus is higher than the one estimated for a beam scanning system analyzed in Sec.~\ref{sec:analysis_scanning}.
\section{Dose rate estimation}
\label{sec:dose_rate_estimation}
Results of the Monte Carlo calculations mentioned above, in conjunction with known parameters of the {\liliput}  accelerator, allows for estimation of the expected dose rate, $\dot{D}$, at the irradiation field.
The parameters used in the estimation are summarized in Table~\ref{tab:parameters}.
From the data in Table~\ref{tab:parameters}, one can expect beam flux at the irradiation plane to be $9.2\cdot10^{11}$~\mbox{electrons/s$\cdot$cm$^2$}. Multiplying the flux by dose deposited per electron, one can estimate the total dose rate to be about $\dot{D}=17$~kGy/min or 1~MGy/h. The assumed beam current of 50~\mbox{mA/imp} is in fact rather conservative assumption, as the {\liliput} can run stably with twice as high beam current. Upper limit for $\dot{D}$ is therefore about 34~kGy/min or 2~MGy/h.
\begin{table}[!hbt]
\centering
\begin{tabular}{l|c} \hline
\hline
Initial beam current & 50 mA/imp \\
Pulse duration & 4 $\mu$s \\
Pulse repetition & 100 Hz \\
Transmission efficiency & 39 \% \\
\hline
Mean energy, $<$$E$$>$, at the irradiation plane & 7.9 MeV \\
Collisional energy loss in air $S_{col}(<$$E$$>)$ & 1.93 MeV cm$^2$/g \\
Dose deposited per electron in 1 cm$^3$ of air & 3.09$\cdot$10$^{-10}$ Gy \\ 
\hline
\end{tabular}
\caption{\small Summary of the input parameters for the estimation of dose rate at the irradiation plane.} 
\label{tab:parameters}
\end{table}
\begin{figure}[!hbt]
  \centering
	\includegraphics[width=0.5\columnwidth]{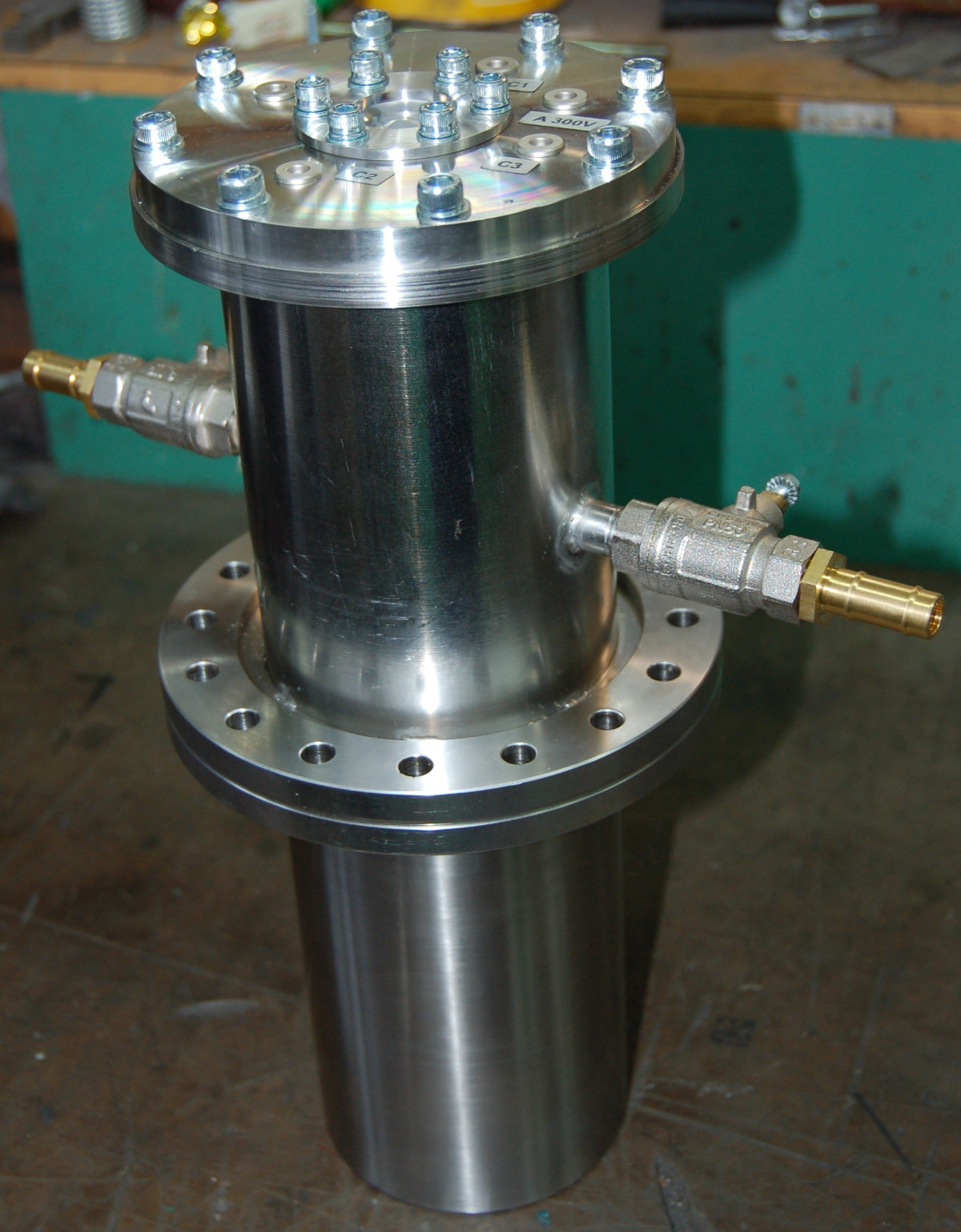}
  \caption{\small Photograph of a part of beam forming system designed and manufactured at NCBJ. Aluminum flattening foil is visible in the center, on top of the steel flange on entry to the applicator tube. A flange visible in the middle of the applicator is for connecting the applicator with the LN$_2$ cryostat, as discussed in Sec.~\ref{sec:requirements}.} 
  \label{fig:Tuba_photo}
\end{figure} 
\section{Conclusions, current state and outlook}
Two possible solutions of the electron beam forming system for high dose rate irradiation at Nondestructive Testing Laboratory at Wroclaw Technology Park were studied. Contrary to expectations, analysis of a simple but realistic model of beam scanning system revealed that this solution would not provide better transport efficiency, and therefore the dose rate, than much simpler dual-foil system. Thus, the latter was chosen for realization. The final system design was optimized based on thorough Monte Carlo modeling of its performance. For the optimized system, the calculated beam transport efficiency amounts to about 39\%. Based on this, the expected dose rate is in range from 17~kGy/min (1~MGy/h) at half the available beam current, up to 34~kGy/min (2~MGy/h) at full available beam current. 
The dual-foil system has been fabricated at the National Centre for Nuclear Research and is scheduled for installation in the Nondestructive Testing Laboratory at WPT in the fall of 2015. The applicator tube with flattening foil attached at its entrance is shown in a photograph in Fig~\ref{fig:Tuba_photo}.

Once installed and commissioned, the electron beam irradiation facility at NDT, apart from serving as a tool for planned research on radiation hard insulating materials, is expected to create prospects for R\&D works in a broad range of potential applications, including radiation hard materials, electronics and semiconductors, enhancement of optical properties of gemstones, hardening and regeneration of coatings, purification of exhaust gases from combustion of fossil fuels, food preservation, etc.


\begin{thebibliography}{99}
\bibitem{Scharf1989} W.~H.~Scharf, \textit{Particle Accelerators - Applications in Technology and Research}, John Wiley \& Sons (1989)
%
\bibitem{Hamm2012} R.~W.~Hamm, M.~E.~Hamm (eds.), \textit{Industrial Accelerators and Their Applications}, World Scientific (2012)
%
\bibitem{AbouMandour1978} M.~Abou~Mandour, D.~Harder, \textit{Strahlentherapie} \textbf{154}, 328 (1978).
%
\bibitem{Grusell1994} E.~Grusell, A.~Montelius, A.~Brahme, G.~Rikner, K.~Russell, \textit{Phys. Med. Biol.} \textbf{39}, 2201 (1994).
%
\bibitem{Kainz2005}  K.~K.~Kainz, J.~A.~Antolak, P.~R.~Almond, C.~D.~Bloch, K.~R.~Hogstrom, \textit{Phys. Med. Biol.} \textbf{50}, 755 (2005).
%
\bibitem{Adrich2014} P.~Adrich \textit{et al}., \textit{Int. J. Mod. Phys. Conf. Ser.} \textbf{27}, 1460125 (2014).
%
\bibitem{Agostinelli2003} S.~Agostinelli \textit{et al}., \textit{Nucl. Instrum. Methods} \textbf{A506}, 250 (2003).
%
\bibitem{ICRU1984} ICRU International Commission on Radiation Units and Measurements, \textit{ICRU Report 37, Stopping Powers for Electrons and Positrons}, (1984).
%
\bibitem{Adrich2015} P.~Adrich, accepted for publication in Nucl. Instrum. Methods Phys. Res. A, available online at http://dx.doi.org/10.1016/j.nima.2016.01.043 
\end{thebibliography}
\end{document}